\documentclass[a4paper]{jpconf}
\usepackage{graphicx}                                  
\usepackage{epstopdf}                                  
\usepackage{bm}                                        
\usepackage{amssymb}                                   
\usepackage{amsmath}                                   
\usepackage{hyperref}                                  

\begin{document}
	
	\title{Self-consistent and Maxwell approximations to describe the excess conductivity anisotropy in FeSe above superconducting transition temperature}
	\author{K K Kesharpu$^{1 \: \dagger}$, P D Grigoriev$^{1\:2\:3}$, D I Lazeva$^{1}$ and T I Mogilyuk$^{4}$}
	
	\address{$^{1}$National University of Science and Technology ``MISiS'', 119049 Moscow, Russia}
	\address{$^2$L. D. Landau Institute for Theoretical Physics, 142432 Chernogolovka, Russia}
	\address{$^3$P. N. Lebedev Physical Institute, RAS, 119991 Moscow, Russia}
	\address{$^4$National Research Center ``Kurchatov Institute", 123098 Moscow, Russia}
	
	\eads{$^{\dagger}$m1708377@edu.misis.ru}

	\begin{abstract}
			Using the self-consistent approximation we calculate conductivity in an anisotropic heterogeneous media with superconducting inclusions and compare the results with those obtained previously using the Maxwell approximation and with available experimental data on excess conductivity above $ T_c= $~8K in FeSe. Advantages and drawbacks of these two approximations are discussed. The obtained analytical formulas can be applied to various other anisotropic heterogeneous superconductors, including high-Tc layered superconductors.
	\end{abstract}
	
	
	\section{Introduction}
	
	Increasing the transition temperature ($T_c$) in high-$ T_c $ superconducting materials is one of the major challenges in condensed matter physics. Most of high-$ T_c $ superconductors are usually metal alloys or complex oxide ceramics. The copper oxides \cite{Keimer2015} and iron based superconductors \cite{Si2016} are of special interests because $T_c$ in these materials is highest at ambient pressure and can be controlled in some range by chemical composition and doping level. During the doping process, the dopant does not deposit uniformly, which transforms the initial spatially homogeneous compound to a heterogeneous material. Because of this inhomogeneity, superconducting (SC) properties are often observed first in spatially disconnected parts of the materials at $T>T_c$ \cite{Kresin2006}. As the temperature $T$ approaches $T_c$, small superconducting areas become larger and their phases become coherent. In the end, at $T=T_c$, the whole volume becomes superconducting. This type of inhomogeneous development of superconducting properties above $T_c$ was called \textit{Gossamer Superconductivity} \cite{Sinchenko2017}. It may also be named \textit{Heterogeneous Superconductivity}. Existence of such heterogeneous phase has been corroborated by magnetic response experiments using \textit{scanning SQUID microscopy} method \cite{Iguchi2001}. 
	Spatial inhomogeneity in superconductor may increase $T_c$, as shown theoretically \cite{Kresin2006, Martin2005}. It is also responsible for superconducting effects above $T_c$. The understanding of relation between spatial inhomogeneity and high-Tc superconductivity is very important for synthesizing new materials with higher $T_c$. 
	
	Recently it was proposed that the excess conductivity of a layered anisotropic heterogeneous compound due to isolated superconducting islands is also very anisotropic with the maximal effect along the least conducting direction \cite{Sinchenko2017,Grigoriev2017}. The quantitative description of this effect using the \textit{Maxwell approximation (MA)} was proposed and used to analyze experimental data on conductivity and diamagnetic response in FeSe \cite{Sinchenko2017,Grigoriev2017}. In the present work we use \textit{self-consistent approximation (SCA)} (See Refs. \cite{Landauer1978,Torquato2002} for short review) to derive analytical expression for conductivity and its anisotropy in such a heterogeneous material with two different phases. Then we compare the results obtained using SCA and Maxwell approximations with the experimental data in bulk FeSe. Finally, we discuss the advantages and drawbacks of these two approximations. 
			
	FeSe has the simplest chemical composition among the large family of iron-based high-Tc superconductors, but its electronic structure and properties are very interesting. Under pressure its transition temperature raises to 40~K \cite{Medvedev09}. Very promising is the observation of superconductivity with $T_c>100$~K in FeSe monolayer on SrTiO$_3$ substrate \cite{GeFeSe}. The bulk FeSe has inhomogeneous microscopic structure and $T_c$ \cite{Naidyuk}. This makes FeSe very convenient for the comparison with our model of anisotropic heterogeneous conductor. 
	
	
	\section{Theory}
	
\begin{figure}[h]
	\centering
	\includegraphics[width=20pc]{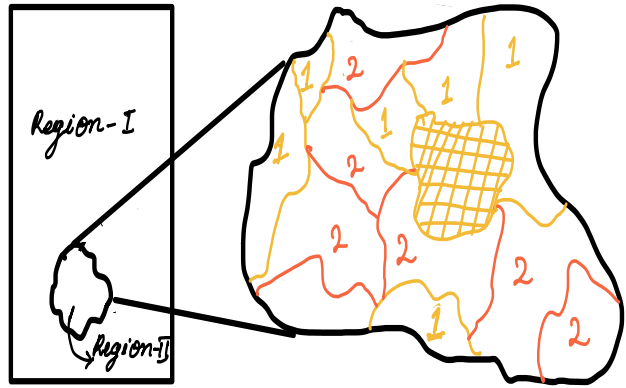}
	\caption[short text]{Square bounded region is the full sample containing region-I ($\textit{Conductivity}\rightarrow \sigma_e$; $\textit{Electric field} \rightarrow E_\text{I}$) and inside it, the embedded small region-II. The blow out of region-II contains phase-1 ($\textit{Conductivity}\rightarrow \sigma_1$; $\textit{Electric field} \rightarrow E_\text{II}$; $\textit{Color} \rightarrow \textit{Yellow}$; $ \textit{Notation} \rightarrow 1$) denoted as notation \textit{1} and phase-2 ($\textit{Conductivity}\rightarrow \sigma_2$; $\textit{Electric field} \rightarrow E_\text{II}$; $\textit{Color} \rightarrow \textit{Red}$; $ \textit{Notation} \rightarrow 2$). The shaded region is of phase-1 surrounded by arbitrary phase.}
	\label{plot_illustration}
\end{figure}

	Conductivity of inhomogeneous media under SCA is found by making 3 primary assumptions: (i) Volume fractions of different phases are distributed randomly, (ii) spatial average of the effective field is equal to the applied field $ E $, (iii) all the included phases are symmetrically treated. For simplicity we will derive the relation for a sample containing only two phases. Let us imagine a sample of heterogeneous material containing two regions (region-I and region-II as shown in Fig. \ref{plot_illustration}). Let the conductivity of region-I, which is completely homogeneous, be $\sigma_e$. Contrary to region-I, region-II is inhomogeneous and contains two phases, i.e. phase-I and phase-II, with conductivity $\sigma_1$ and $\sigma_2$ respectively. Since both phases in region-II are treated symmetrically, the conductivity outside the boundary of a phase does not depend on which phase inclusion is considered. For example, conductivity outside the shaded part in region-II should be the same (\textit{i.e., effective conductivity $(\sigma_e)$}), whether the shaded part belongs to phase-I or phase-II. This idea is incorporated in the mathematical derivation of $\sigma_e$ by thinking that islands of conductivity $\sigma_1$ or $\sigma_2$ are embedded  inside a large homogeneous material of conductivity $\sigma_e$.
	
	Let $E_\text{I}$ and $E_\text{II}$ be the electric field inside regions I and II respectively on Fig. \ref{plot_illustration}. These two electric fields should be equal, $E_\text{I}=E_\text{II}$, because SCA assumes that the effective conductivity value across the material on average is the same, and that the current flowing in both regions I and II is also the same, $j_\text{I} = j_\text{II}$. These two assumptions, combined with the relation $j=\sigma E$, imply that $E_\text{I}=E_\text{II}$. 
	First we take spherical islands of different phases. Then the dipole moment of $n_1$ spherical islands of radius \textit{$a_1$} and of conductivity $\sigma_1$ embedded inside a material of conductivity $\sigma_e$ can be written as
	\begin{equation}
	P_1 = \frac{n_1 E_\text{II} 4 \pi a_1^3 (\sigma_ 1 - \sigma_e)}{3(\sigma_ 1 + 2 \sigma_e)} =  \frac{\phi_1 E_\text{II} (\sigma_ 1 - \sigma_e)}{(\sigma_ 1 + 2 \sigma_e)} \:  .
	\label{eqn_dipole_moment_1}
	\end{equation}
	Similarly the dipole moment of $n_2$ spherical islands of radius \textit{$a_2$} and of conductivity $\sigma_2$ embedded inside a material of conductivity $\sigma_e$ is given by 
	\begin{equation}
	P_2 = \frac{n_2 E_\text{II} 4 \pi a_2^3 (\sigma_ 2 - \sigma_e)}{3(\sigma_ 2 + 2 \sigma_e)} =  
	\frac{\phi_2 E_\text{II} (\sigma_ 2 - \sigma_e)}{(\sigma_ 2 + 2 \sigma_e)} \:  .
\label{eqn_dipole_moment_2}
	\end{equation}
	 
	 The presence of islands of both phases inside region-II creates a polarization which disturbs the electric field locally. But on average their contributions have to cancel each other to satisfy the condition $E_\text{I}=E_\text{II}$, and the total polarization of region-II is zero. Mathematically it can be written as $P_1 +P_2 = 0$,
	which with the help of Eqs. (\ref{eqn_dipole_moment_1}) and (\ref{eqn_dipole_moment_2}) gives the relation for $\sigma_e$:
	\begin{equation}\label{eqn_spherical_inclusion}
		\frac{\phi_1 (\sigma_ 1 - \sigma_e)}{(\sigma_ 1 + 2 \sigma_e)} + \frac{\phi_2 (\sigma_ 2 - \sigma_e)}{(\sigma_ 2 + 2 \sigma_e)} =0\:.
	\end{equation}
For superconducting inclusion inside a metallic phase we need to substitute $\sigma_1 \rightarrow \sigma_s$; $ \sigma_2 \rightarrow \sigma_m$; $ \phi_1 \rightarrow \phi_{s}$; $ \phi_2 \rightarrow \phi_{m}$ in 
Eq. (\ref{eqn_spherical_inclusion}).

	For the islands of ellipsoidal shape instead of single depolarization value, as in spherical case, we have 3 depolarization values for 3 main axes of the ellipsoid. We consider prolate spheroidal inclusions with $a_x=a_y$, $ a^2_z/a^2_x \equiv \gamma > 1$ and the eccentricity $\chi \equiv \sqrt{1-1/\gamma}$.  After algebraic operation with depolarization matrix (See Eq. (18.19) of Ref. \cite{Torquato2002} for the whole procedure) the equation for effective conductivity $\sigma_e$ can be written as
	\begin{equation}
	\label{eqn_condition_ellipsoidal_inclusion}
		\phi_s(\boldsymbol{\sigma_s} - \boldsymbol{\sigma_e}) R^{se} + \phi_m(\boldsymbol{\sigma_m} - \boldsymbol{\sigma_e}) R^{me} = 0 \:  ,
	\end{equation}
	where $R^{je} = \left[ I + A^* \boldsymbol{\sigma_e}^{-1} (\boldsymbol{\sigma_j} - \boldsymbol{\sigma_e})\right]^{-1}$, $I$ is the identity matrix, $\sigma_j$ is the conductivity in phase $j$, $\boldsymbol{\sigma_j}=I\sigma_j$, and the effective conductivity tensor $\boldsymbol{\sigma_e}$ is diagonal:
	\begin{equation}\label{eqn_sigma_e_matrix}
	\boldsymbol{\sigma_e} =
	\begin{bmatrix}
	\sigma_{xx} & 0 & 0 \\
	0 & \sigma_{yy} & 0 \\
	0 & 0 & \sigma_{zz}
	\end{bmatrix} \:  .
	\end{equation}
 The depolarization tensor $A^*$ for a prolate ellipsoid is given by (see section 17.1.2 of Ref. \cite{Torquato2002})
	\begin{equation}\label{eqn_polarizability_matrix}
	A^*=
		\begin{bmatrix}
		Q & 0 & 0 \\
		0 & Q & 0 \\
		0 & 0 & 1-2Q
		\end{bmatrix} \:  ,
	\end{equation}
	where
	\begin{equation}\label{eqn_value_of_Q}
		2Q = 1 + \frac{1}{\gamma - 1} \left[ 1 - \frac{1}{2 \chi} \ln \left(\frac{1+\chi}{1-\chi}\right)\right] \:  .
	\end{equation}
	We substitute the values of $\sigma_m$, $A^*$, $\sigma_e$ and $\phi_m=(1-\phi_s)$ in Eq. (\ref{eqn_condition_ellipsoidal_inclusion}), and simplify using that in the superconducting phase $\sigma_s \to \infty $. Then solving for $\sigma_e$ we get
	\begin{equation}\label{eqn_effective_conductivity_ellipsoidal}
	\frac{\sigma_{xx}}{\sigma^x_m}=	\frac{\sigma_{yy}}{\sigma^y_m} = \frac{Q}{Q-\phi_s}\: ,  \:  \frac{\sigma_{zz}}{\sigma^z_m} = \frac{1-2Q}{1-2Q-\phi_s}  \:  ,
	\end{equation}
	which is valid only for the isotropic metallic conductivity $\sigma^x_m=\sigma^y_m=\sigma^z_m$.
	
	The anisotropic case, when $\sigma^x_m\neq \sigma^y_m\neq \sigma^z_m$, can be transformed to isotropic by dilation with different coefficient along different axes. For layered compounds with in-plane isotropy $\sigma^x_m= \sigma^y_m > \sigma^z_m$, similar the procedure in Refs. \cite{Sinchenko2017,Grigoriev2017}, we dilate only along the z-axis as $z^* = z/\sqrt{\eta}$, where $\eta=\sigma^z_m/\sigma^x_m$, so that after dilatation $\sigma^x_m=\sigma^y_m=\sigma^z_m$. The effective conductivity for this case can be obtained from Eq. (\ref{eqn_effective_conductivity_ellipsoidal}) by making the substitution $\gamma \rightarrow \gamma^* \equiv \gamma/\eta$. For the highly anisotropic case, i.e when $\sigma_{zz} \ll \sigma_{xx}$ and $\gamma^*\gg 1$, Eq. (\ref{eqn_value_of_Q}) simplifies to
	\begin{equation}\label{eqn_modified_value_of_Q}
		Q = \frac{1}{2} + \frac{2-\ln(4 \gamma^*)}{4\gamma^*} \:  .
	\end{equation} 
	Then for prolate spheroid with $(a_z \gg a_x =a_y)$ Eq. (\ref{eqn_effective_conductivity_ellipsoidal}) gives 
		\begin{equation}\label{eqn_modified_effective_conductivity_ellipsoidal}
			\begin{aligned}
			&\frac{\sigma_{xx}}{\sigma^x_m} = \frac{2\gamma^* + 2 - \ln (4\gamma^*)}{2\gamma^* + 2 - \ln (4\gamma^* )- 4\gamma^*\phi_s} \ ,\\
			&\frac{\sigma_{zz}}{\sigma_m^z} = \frac{ \ln (4\gamma^*) -2}{ \ln (4\gamma^*) -2- 2 \gamma^* \phi_s} \ ,
			\end{aligned}
		\end{equation} 
which remains valid also after the inverse dilatation to the original geometry.	The value of $\gamma=a_z^2/a_x^2$ is determined by the shape of spheroids and can be taken arbitrarily, provided $\gamma^* = (a_z^2 \sigma_{xx})/(a_x^2 \sigma_{zz})>1$ (for $\gamma^*<1$ one should use Eq. (17.31) of Ref. \cite{Torquato2002} instead of Eq. (\ref{eqn_value_of_Q})). Eq. (\ref{eqn_modified_effective_conductivity_ellipsoidal}) differs strongly from the expression for conductivity in such a heterogeneous system obtained previously using the Maxwell approximation\cite{Sinchenko2017,Grigoriev2017}:
\begin{equation}
\frac{\sigma _{xx}}{\sigma^x_m}\approx \frac{1}{1-\phi_s }+\phi_s ,\
 \frac{\sigma _{zz}}{\sigma_m^z}\approx \frac{1}{1-\phi_s }+\frac{%
	2\gamma^* \phi_s }{\ln \left( 4\gamma^* \right) -2}. \label{sSCf}
\end{equation}
In particular, conductivity in Eqs. (\ref{eqn_modified_effective_conductivity_ellipsoidal}) and (\ref{sSCf}) diverges at completely different $\phi_s$ values.
	
 	\section{Comparison with experiment and with Maxwell approximation}
 	
 	In this section we compare the above theoretical calculation with the experimental data from 
 	Refs. \cite{Sinchenko2017}  and \cite{Grigoriev2017} on resistivity in FeSe along the conducting layers \textit{a-b} ($\rho_{ab}$) and perpendicular to them ($\rho_c$). There are several reasons for choosing FeSe. First, FeSe is a heterogeneous\cite{Naidyuk} quasi-2D superconductor whith large conductivity anisotropy: its resistivity along the \textit{z} axis is about 400 times greater than along the  conducting \textit{x-y} planes, but the in-plane conductivity $\sigma_{ab}$ is isotropic. Hence, FeSe is a perfect system to apply the prolate spheroid case as described in the previous section. Second, the excess conductivity above $T_c$ due to superconducting inclusions in the Maxwell approximations has already been calculated, and the corresponding volume ratio ($\phi_s$) of superconducting phase for the same system has already been extracted, which allows us to compare both the results. The $\phi_s$ values from Eq. (\ref{eqn_modified_effective_conductivity_ellipsoidal}) can be found as 
 	\begin{equation}\label{eqn_SC_island_ratio_ellipsoidal}
 	\begin{aligned}
 	&\phi_s^x =\frac{(\sigma_{xx}-\sigma_m^z)Q}{\sigma_{xx}}\approx \frac{(\sigma_{xx}-\sigma_m^x)\left[2\gamma^* + 2-\ln(4\gamma^*)\right]}{4 \gamma^*\sigma_{xx}}\ , \\
	&\phi_s^z=\frac{(\sigma_{zz}-\sigma_m^z)(1-2Q)}{\sigma_{zz}}\approx\frac{(\sigma_{zz}-\sigma_m^z)\left[\ln(4\gamma^*)-2\right]}{2\gamma^*\sigma_{zz}} \ ,
 	\end{aligned}
 	\end{equation} 
 	where $\phi_s^x$ and $\phi_s^z$ are the superconducting volume ratios calculated from experimental conductivity along the $ x $ and $ z $ axes respectively. In Ref. \cite{Grigoriev2017} it was shown experimentally that conductivity below $ T=40-50 $~K has starts to deviate from the linear temperature dependence, hence we suspect that the SC islands starts to appear around $ T=40 $~K. We find the metallic conductivities ($\sigma_m^x,\sigma_m^z$) by taking the inverse of linear extrapolated resistivity from high temperatures. In FeSe the conductivity along \textit{z}-axis is around 400 less than along \textit{x-y} plane for temperature below 50K. Hence, we use the value $\eta \approx 1/400$ for FeSe to calculate $\gamma^* \equiv \gamma / \eta$.
 	
 	\begin{figure*}[h]
 		\centering
 		\includegraphics[width=20pc]{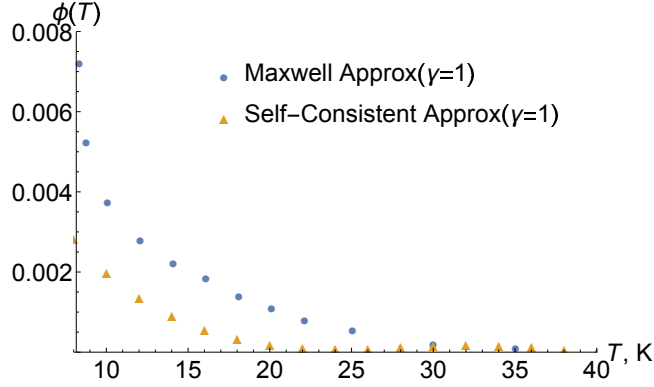}
 		\caption[short text]{The volume ratio ($\phi_s^z$) extracted from experimental data on resistivity from Ref. \cite{Sinchenko2017} using Eq. (\ref{eqn_SC_island_ratio_ellipsoidal}) under self-consistent approximation (orange triangles) and extracted from Fig. 4 of Ref. \cite{Grigoriev2017} under Maxwell approximation (blue circles). We used the value $\gamma =1$ (spherical islands) and anisotropy ratio $\eta =\sigma^z_m/\sigma^x_m = 1/400$.}
 		\label{plot_volume_ratio}
 	\end{figure*}
 
	\begin{figure*}[h]
  	\begin{minipage}{18pc}
 	\includegraphics[width=18pc]{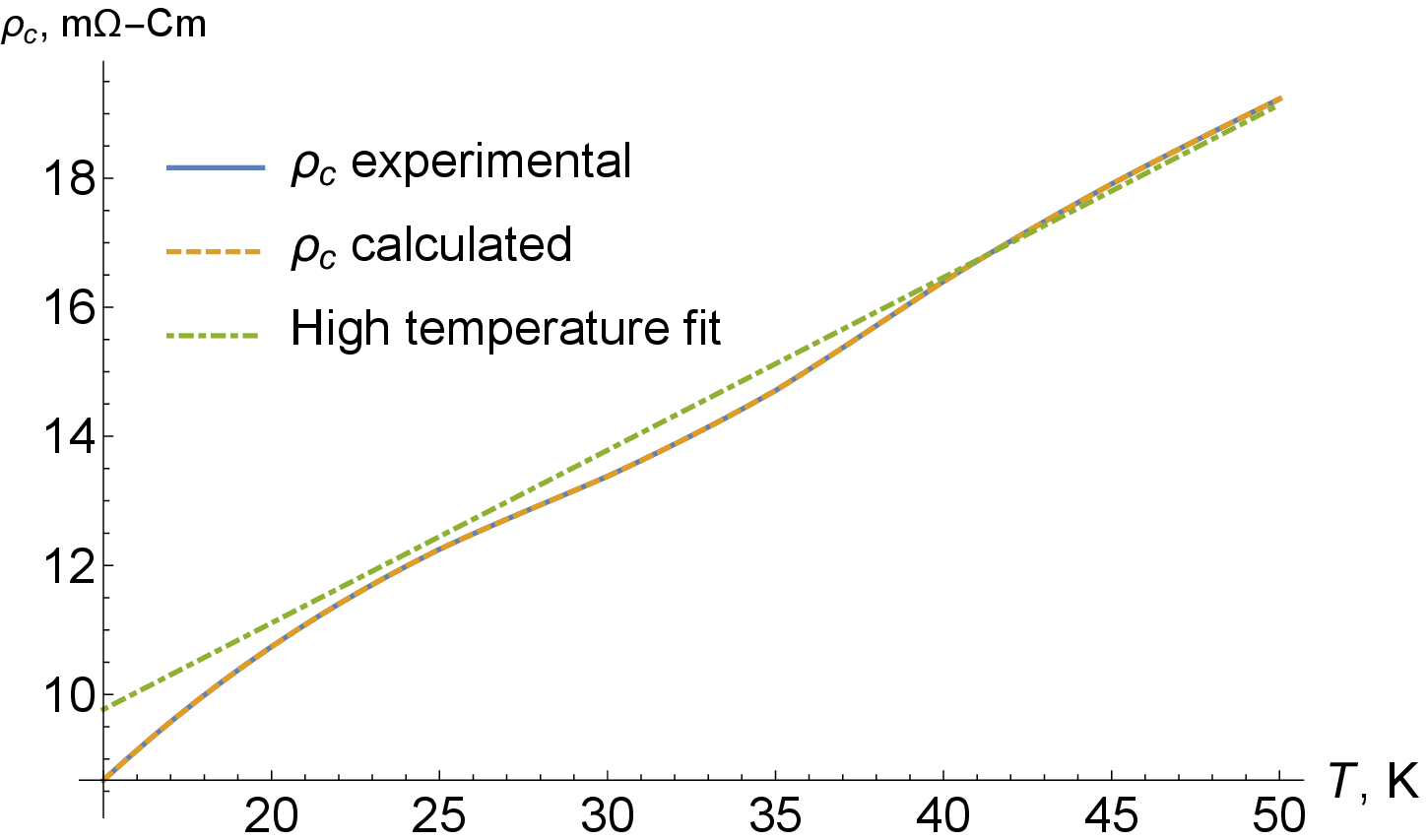}
 	\caption{Comparison of experimental interlayer resistivity from Fig. 2c of Ref. \cite{Sinchenko2017} and theoretical result in Eq. (\ref{eqn_modified_effective_conductivity_ellipsoidal}). High temperature fit $\rho_c = 6.5 +0.3T \text{(m}\Omega\cdot\text{cm)}$ is found by extrapolating $\rho_c$ above 50~K. }
 	\label{plot_rho_c(T)}
 	\end{minipage}\hspace{2pc}
	\begin{minipage}{18pc}
 	\includegraphics[width=18pc]{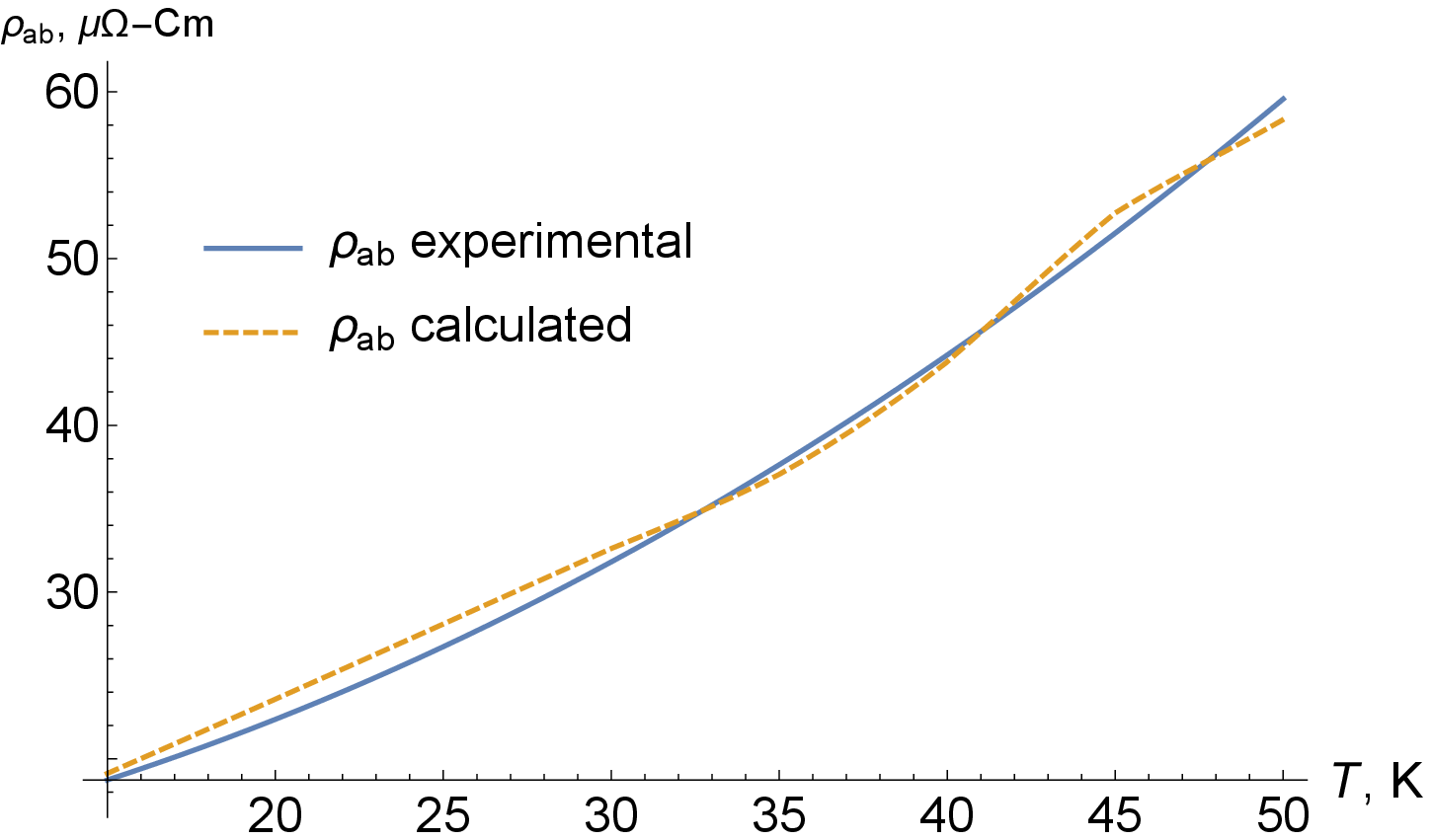}
 	\caption{Comparison of experimental in-plane resistivity from Fig. 2c of Ref. \cite{Sinchenko2017} with theoretical result in Eq. (\ref{eqn_modified_effective_conductivity_ellipsoidal}). High temperature fit $\rho_{ab}=12.3 +0.21 T (\mu \Omega \cdot\text{cm)}$ is found by extrapolating above 50~K.}
 	\label{plot_rho_ab(T)}
 	\end{minipage}
 	\end{figure*}
 
 	We use the conductivity along z-axis to find $\phi_s^z$ and then use $\phi_s^z$ to calculate conductivity along the xy plane. We choose $ z $-axis to find the volume ratio $\phi$ because the effect of SC islands is most significant in this direction, as it is the least conducting axis.	
 	In Fig. \ref{plot_volume_ratio} we plot the superconducting volume ratio calculated using Eq. (\ref{eqn_SC_island_ratio_ellipsoidal}). We extracted resistivity data from Fig.~2c of Ref. \cite{Sinchenko2017}. Metallic conductivity $\sigma_m^z =1/[ 0.0065 +0.0003T] \text{(m}\Omega\cdot\text{cm)}^{-1}$ is found by extrapolating the $\rho_c$ from 50~K upwards. The hump in the volume ratio is the due to decrease in the slope of $\rho_c$ around temperature of $30<T<35$. For comparison, the SC volume fraction $\phi(T)$ extracted from Fig.~4 of Ref. \cite{Grigoriev2017} using Maxwell approximation for $\gamma=1$ is also plotted in Fig. \ref{plot_volume_ratio}. Here we should mention that as we don't have exact metallic resistivity, we have made the linear approximation for $\rho_c$ which might include some error of small percentage. In Figs.~\ref{plot_rho_ab(T)} and \ref{plot_rho_c(T)} the comparison of experimental and theoretical resistivity is shown. The complete agreement of prediction and experiment in Fig. \ref{plot_rho_c(T)} is because we calculated the $\phi_s$ using the same resistivity along the $ z $ axis.
 	
 	\section{Discussion}
 	
 	Above we derived the analytical expressions for conductivity in the heterogeneous superconductor using the  self-consistent approximation (SCA) and compare the results with those\cite{Sinchenko2017,Grigoriev2017} obtained using Maxwell approximation and with experimental data in FeSe. The Maxwell and self-consistent approximations are the simplest to describe conductivity in heterogeneous materials \cite{Torquato2002}. They allow obtaining analytical results convenient for physical predictions and for comparison with experiment. The more accurate methods do not give simple analytical expressions and require the knowledge of spatial and size distribution of the second-phase inclusions \cite{Torquato2002}. The limitations and drawbacks of the SCA and Maxwell approximations differ, therefore their comparison is helpful for a reliable description of real materials. 
 	
 	The Maxwell approximation is valid only at small volume fraction $\phi \ll 1$ of the second phase, i.e. of superconducting inclusions in our case, and it gives incorrect percolation threshold $\phi =1$. However, it is strictly substantiated in the limit $\phi \ll 1$ and coincides with the lower bound for the effective conductivity of media with superconducting inclusions at arbitrary $\phi $. The SCA sometimes works well even at $\phi \sim 1$ and gives almost correct percolation thresholds in isotropic 2D and 3D cases. However, in anisotropic case these
 	percolation thresholds differ for different directions, which is incorrect for most superconductors. Thus, in Eqs. (\ref{eqn_effective_conductivity_ellipsoidal}) and (\ref{eqn_modified_effective_conductivity_ellipsoidal}) the conductivity along $ z $ and $ x $ axes diverges at different values of $\phi_s$. The stronger is anisotropy of metallic phase, the larger is the difference between percolation thresholds. This is generally incorrect, but it may somewhat reflect physical situation if the length of SC islands or clusters is comparable to the size of the whole sample. In real superconductors, taking into account the Josephson coupling and proximity effect, the effects of anisotropic percolation thresholds may also appear, but in a different way.  
 	Thus, it is not a priori clear whether the SCA or Maxwell approximation is better for strongly anisotropic heterogeneous conductors. However, one may say that while the Maxwell approximation gives the lower bound, the SCA gives the upper bound for the effective conductivity along the least conducting axis. It is reflected in Fig. \ref{plot_volume_ratio}, where the same excess conductivity in the SCA is obtained for much smaller volume fraction of superconducting inclusions than in the Maxwell approximation. Hence, both these approximations are helpful in analyzing experimental data.
 	
 	\section{Acknowledgments}
	The work was carried out with the financial support from the Ministry of Education and Science of the Russian Federation in the framework of increase Competitiveness Program of NUST ``MISIS'', implemented by a governmental decree dated 16th of March 2013, No 211, and from the ``Basis'' Foundation. Sec. III was supported by the RSF grant \# 16-42-01100. P.G. thanks RFBR grant \#18-02-00280. T.M. thanks RFBR grant \# 18-02-01022.

\end{document}